\documentclass[prb,amsmath,amssymb,showkeys]{revtex4}

\usepackage{graphicx}
\usepackage{dcolumn}
\usepackage{bm}
\usepackage{graphicx}
\usepackage{amsmath}
\usepackage{enumerate}
\usepackage{comment}
\usepackage{color}
\begin{document}

\title{On the Stability of Community Detection Algorithms on
Longitudinal Citation Data}

\author{Michael J. Bommarito II (mjbommar@umich.edu)}
	\affiliation{Department of Political Science, University of Michigan, Ann Arbor}
	\affiliation{Department of Mathematics, University of Michigan, Ann Arbor}
	\affiliation{Center for the Study of Complex Systems, University of Michigan, Ann Arbor}

\author{Daniel Katz (dmartink@umich.edu)}
	\affiliation{Department of Political Science, University of Michigan, Ann Arbor}
	\affiliation{Gerald R. Ford School of Public Policy, University of Michigan, Ann Arbor}
	\affiliation{Center for the Study of Complex Systems, University of Michigan, Ann Arbor}

\author{Jon Zelner (jzelner@umich.edu)}
	\affiliation{Department of Sociology, University of Michigan, Ann Arbor}
	\affiliation{Gerald R. Ford School of Public Policy, University of Michigan, Ann Arbor}
	\affiliation{Center for the Study of Complex Systems, University of Michigan, Ann Arbor}

\date{August 1, 2009 (updated August 17, 2009)}

\begin{abstract}
\textbf{Abstract:} There are fundamental differences between citation networks
and other classes of graphs.  In particular, given that citation networks are
directed and acyclic, methods developed primarily for use with undirected social
network data may face obstacles.  This is particularly true for the dynamic
development of community structure in citation networks.  Namely, it is neither
clear when it is appropriate to employ existing community detection approaches
nor is it clear how to choose among existing approaches. Using simulated data,
we attempt to clarify the conditions under which one should use existing methods
and which of
these algorithms is appropriate in a given context. We hope this paper will
serve as both a useful guidepost and an encouragement to those interested in the
development of more targeted approaches for use with longitudinal citation data.
\end{abstract}

\keywords{community detection, clustering, network dynamics, citation
networks, directed acyclic graphs}

\maketitle
\pagebreak

\section{Introduction \& Motivation}
We live in age of data.  Never before has so much wide spread information been
available to an interested researcher. Advances in computing power and decreases
in the cost of data storage have made available for analysis larger and larger
streams of information on social, economic and political systems.   Scholars
have taken notice, with some leading figures calling for a new age of
computational social science (Lazer, et al. 2009\cite{Lazer2009}). Indeed, while
the structure of science generally involves the entrenchment of status quo
practices, there is mounting evidence that computing is fundamentally altering
the nature of scholarly inquiry.

In seeking to contribute to the cause, the focus of this inquiry is the dynamic
analysis of longitudinal citation data.  The study of citation patterns has a
long history with pioneering work in not only the social and physical
sciences but also interdisciplinary fields such as bibliometrics and
scientometrics. Despite its early importance and significant potential, the
analysis of citations has not been as wide spread as one might imagine. 
However, this trend appears to be reversing with literatures in a variety of
disciplines indicating recently renewed interest (e.g., Bommarito, et al. 2009
\cite{Bommarito2009}, Choi and Gulati 2008\cite{Choi2008}, Wright and Armstrong
2008\cite{Wright2008}, Fowler, et al. 2007\cite{Fowler2007}, Leicht 2007, et
al.\cite{Leicht2007}, Lehmann, et al. 2003\cite{Lehmann2003}).
\par
There are a wide variety of analytic techniques available to study longitudinal
citation data including econometric and network-based approaches.
Focusing upon the latter, we believe the evolution of longitudinal citation
networks may offer tremendous insights for a given substantive question. Whether
tracing the spread of ideas through academic citation networks, the development
of common law systems or following the spread of innovation in the patent
database, understanding the dynamics contained within such representations can
help enrich substantive theories of continuity and change within social,
economic and political systems. 
\par
It is important to note that such data can and should be considered at many
levels - from microscopic node attributes to macroscopic characterizations at
graph level. On the spectrum of resolution between these two extremes are the
mesoscopic structures the literature commonly characterized as communities.
Although the precise definition of a community remains a bit elusive, Porter,
Onnela and Mucha\cite{Porter2009} note communities ``consist of a group of nodes
that are relatively densely connected to each other but sparsely connected to
other dense groups in the network.''
\par
Developing strategies designed to identify and better understand the patterns
characteristic of these mesoscopic structures has consumed the attention of many
in the networks science literature. For very small networks, it is possible to
visually identify and explain the development of these subgroups. However, in
most cases, it is necessary to employ algorithmic techniques designed to
identify the cohesive subgroups of a given network. Recent years witnessed the
production of a wide variety of useful automated community detection methods.
Algorithms developed by Mark Newman have been signiﬁcantly embraced by the
broader applied literature. In particular, much of the existing applied work
employs approaches such as fast-greedy modularity (Newman 2004
\cite{Newman2004}), leading eigenvector (Newman 2006\cite{Newman2006}) and
edge-betweenness (Girvan and Newman 2002\cite{Girvan2002}). Taken together,
these methods contributed signiﬁcantly to the development of a wide class of
substantive theory. Given these algorithms are familiar to many in the field, we
refer those interested in detailed descriptions to review articles such as
Porter, Onnela and Mucha\cite{Porter2009}, Fortunato and Castellano
\cite{Fortunato2009}, or Danon, et al. \cite{Danon2005}.
\par
While undirected social networks have been a primary focus of the literature on
community detection, such methods can be applied to other forms of graphs
including citation networks. Although citation networks display features
consistent with other types of graphs, there exist fundamental differences that
must be confronted by the researcher. In particular, given that citation
networks are directed and acyclic, methods developed primarily for use with
social network data may face obstacles (Leicht, et al. 2007\cite{Leicht2007}).
In particular, when one considers the dynamic development of community structure
in citation networks, it is neither clear when it is appropriate to employ
existing community detection approaches nor is it clear how to choose among
existing approaches. Using simulated data, our analysis attempts to clarify the
conditions under which one should use existing methods and which of these
algorithms is appropriate in a given context. We hope this paper will serve as
both a useful guidepost and an encouragement to those interested in the
development of more targeted approaches for use with longitudinal citation data.
\par

\section{Properties of Citation Graphs}
Citation networks are inherently directed graphs.  The \textit{citing} node asserts that a relationship exists with the \textit{cited} node, never vice versa. Citation networks are the result of a generative process with identifiable constraints: arcs are only created when the tail node of the arc is created, and the head nodes must exist prior to the tail node.  In other words, documents can only cite documents that they have observed on their filtration $\mathcal{F}$.  This constraint, however, can be interpreted both through the lens of the generative process as well as through the lens of the resulting data.  In the discrete-time generative framework, this constraint is equivalent to defining the probability space of arcs from which new arcs are sampled at time $t$:
\begin{align}
	\Omega_t^{A} =& \{(x,y) : x \notin V(G_{t-1}), y \in V(G_{t-1})\}
\end{align}
where $G_{t-1}$ is the time-indexed graph at time $t-1$, and thus $V(G_{t-1})$ is the set of nodes at the previous time step.
\par

From the statistical framework, in which only the resulting graph $G$ of the generative process is observed, we can formalize this constraint as
\begin{align}
	T(x) \leq  T(y) \Leftrightarrow \mathbb{P}((x,y) \in A(G)) = 0
\end{align}
where $T(x)$ is the arrival time of node $x$ and we assume no two events are simultaneous.  The weak inequality on the left-hand side takes into account the case in which $x=y$, thus ensuring that there are no loops in the graph.
\par
These constraints can be shown to imply an even stronger property of the
citation network $G$.  Given a node $x$, there is no directed path from $x$ to
itself.  To see this, examine the simplest case, in which $x$ and $y$ are nodes
with $T(x) < T(y)$.  There are two ways in which a path $(x,x)$ could exist: $x
\rightarrow x$ and $x \rightarrow y \rightarrow x$.  However, as can be seen by
condition (2) above, $T(x) \leq T(x) \leq T(y)$, and thus neither $x \rightarrow
x$ nor $y \rightarrow x$ can exist.  This property is equivalent to stating that
the graph has no cycles of any order - that is, that $G$ is acyclic
\footnote{Preprint repositories such as the arXiv or SSRN allow authors to cite
each other prior to final publication, thus introducing the possibility of
cycles.  However, considering preprints and final publications as separate
nodes corrects for this problem.  Furthermore, considering the full set of
citations in a given field, these citations are unlikely to represent a
significant proportion.}.
\par
These facts clearly indicate that $G$ is in the set of directed acyclic graphs (DAGs).  Directed acyclic graphs are a well-studied family of graphs, often used to model scheduling in software or hardware or causality in Bayesian modeling (Ahmad, et al. 1996 \cite{Ahmad1996}, Madigan, et al. 1995 \cite{Madigan1995}).  Furthermore, there has been some recent work on deriving the equivalent of an Erdos-Renyi models for directed acyclic graphs (Karrer 2009 \cite{Karrer2009}) and on adapting the linear preferential attachment mechanism to directed acyclic graphs (Bollobas, et al. 2003 \cite{Bollobas2003}).  
\par
Directed acyclic graphs have a number of unique properties.  For example, there exists at least one node with in-degree zero (\textit{source}) and at least one node with out-degree zero (\textit{sink}).  Sinks play a critical conceptual role in citation networks, as they represent the introduction of novel information into the citation network.  In the context of empirical data, sinks may also indicate where concepts are imported from sources outside of the data set.  In addition, every directed acyclic graph has a topological ordering of its nodes such that arcs may only be directed corresponding to one orientation of the ordering.  Conditions (1) and (2) can again be seen to specify a topological ordering of a citation network.  Note that for a general directed acyclic graph, there may be more than one valid topological ordering, and standard topological sorting algorithms may return topological orderings for citation networks that do not respect the underlying time-ordering (Bang-Jensen and Gutin 2000 \cite{BangJensen2000}).

\section{Model}
In order to test the stability of these algorithms, we implement a random model
of citation networks.  The model is a discrete-time system driven by a small
number of simple dynamics.  At time $t=0$, $G_0$ is given by $|V(G_0)| = V_0$
and $A(G_0) = \emptyset$.  Note that every node in $V(G_0)$ is a sink.  However,
for the remainder of this paper, we take $V_0$ to be 1 \footnote{Future work
will investigate the sensitivity of the model to this parameter.}.
\par
Given this initial graph, the number of nodes introduced into the graph is
modeled by a homogeneous Poisson process with rate $\lambda$.  We select a
Poisson process because it is a well-understood counting process whose
stationarity and independence assumptions represent a simple null model.  In
addition, in order to analytically derive various graph properties it is
necessary to rely upon compound Poisson processes.  This is not deeply troubling
because compound Poisson processes are defined as the sum of $N(t)$ many IID
random variables, where $N(t)$ is a Poisson process.  These processes have other
of useful properties.  For example, the expected value of a compound Poisson
process is simply the product of the expectation of the counting process $N(t)$
and the expectation of the IID variables to be summed (Grimmet and Stirzaker
2001 \cite{Grimmet2001}).  As a final point, one can easily imagine systems in
which the rate $\lambda$ can vary over time, and the analysis can thus be
carried out by switching to a non-homogeneous Poisson process.
\par
For each new node $n$, we also need to model the arcs introduced by this node
$n$.  We again choose a homogeneous Poisson process with rate $\mu$ per node, as
the underlying assumptions produce a reasonable null model and the resulting
analytics are tractable.  However, we must still determine the process by which
the receiving node of each of these new arcs is to be determined.  This choice
is at the heart of the Erdos-Renyi and Barabasi-Albert models, the two most
well-known random graph models.  In the case of the Erdos-Renyi $G(n,p)$ model,
the existence of every edge is determined by IID Bernoulli experiments with
probability $p$, and thus the number of edges for a node introduced at time $t$
is distributed binomially with $N=|V(G_t)|, p$.  It is clear that the expected
number of citations per document is thus an increasing function of the number of
nodes in the graph.  Thus, as the number of documents in the graphs grows in
time, so too must the average number of citations per new document.  This
acceleration makes it harder to characterize the growth of the graph, and thus
makes it more difficult to infer diagnostic prescriptions from the model. 
Furthermore, the $G(n,m)$ model suffers from a similar problem and would require
$m$ to vary as a function of $N(t)$ (Bollobas 1985\cite{Bollobas1985}).
\par
The development of the Barabasi-Albert (B-A) model was motivated by the
desire to identify a plausible generative process for networks whose degree
distribution follows the ubiquitous power law distribution (Barabasi 1999
\cite{Barabasi1999}).  The B-A model identifies a particular flavor of
preferential attachment as a capable generative mechanism.  Typical
implementations of preferential attachment focus upon the degree of nodes in an
undirected graph.  When considering graphs with direction, such as citations
graphs, it is necessary to modify the B-A model to capture this feature.  
\par
Consider first the original B-A model: let the initial graph $G_0$ consist of a
single node with no edges.  At the first time step, a second node is added to
the graph and an edge connects the first two nodes with probability 1.  The
third node's addition introduces the first element of stochasticity, as an edge
is created from the third node to one of the first two nodes with equal
probability.  At this point, one of the first two nodes now has degree two,
while the remaining two nodes have only degree one.  Though all nodes still have
a positive probability of receiving new edges, the node with degree two is now
more likely to receive future edges.
\par
When this same example is adapted to in-degree with B-A linear preferential
attachment probability, a number of problems occur.  We again let the initial
\textit{directed} graph $G_0$ be a single node with no edges.  At the first time
step, a second node is again added and an arc connects these first two nodes
with probability 1.  At this point, however, the first node has in-degree one,
while the second node has in-degree zero.  According to the adapted equations,
the second node has probability zero of receiving new edges; in fact, the first
node will receive every new edge.  In general, if this process is begun with an
initial graph of $V_0$ nodes and each new node has $m$ edges, then it can be
shown that only $\min(V_0, m)$ of the initial $V_0$ nodes will ever receive
edges.  This absorbing state is obviously not the system we hope to model.
\par
In order to correct for these issues, we design our own arc probability model.  This model is parameterized to allow for a wide class of systems to be modeled, including systems similar to the Erdos-Renyi and Barabasi-Albert models above.  Recall from condition (1) above that the probability space of arcs is given by
\begin{align*}
	\Omega_t^{A} =& \{(x,y) : x \notin V(G_{t-1}), y \in V(G_{t-1})\}
\end{align*}
\par
To specify our model, we would like to assign a probability measure $\mathbb{P}$ over this set of arcs.  This measure determines how the cited node is chosen for each citation arc, and should thus be able to represent our conception of this dynamic.  To do so, $\mathbb{P}$ should be allowed to incorporate both the in-degree and out-degree of each ``citable'' node.  Furthermore, the measure must allow every ``citable'' node to feature some non-zero probability of being cited, thus preventing the graph from becoming deterministic or near-deterministic.  Based on these goals, we assign the following measure
\begin{align}
	\mathbb{P}((x,y)) =& \frac{\exp(\alpha \delta^-(y) + \beta \delta^+(y))}{\sum_{i=1}^{|V(G_t)|} \exp(\alpha \delta^-_i + \beta \delta^+_i)}
\end{align}
where $\delta^-_i$ and $\delta^+_i$ are the in- and out-degree of the $i^{th}$ node of the graph.  The parameters $\alpha$ and $\beta$ are the weights on in- and out-degree respectively, and control how important each node attribute is for the node's probability of receiving cittaions.  For the remainder of this paper, we set $\beta = 0$ to reduce the number of parameters and focus on the effect of in-degree.  Our conception of the dynamics of citation networks leads us to believe that in-degree is much more important than out-degree, and thus $\mathbb{P}$ thus can be simplified to
\begin{align}
	\mathbb{P}((x,y)) =& \frac{\exp(\alpha \delta^-(y))}{\sum_{i=1}^{|V(G_t)|} \exp(\alpha \delta^-_i)}
\end{align}
\par
Despite our choice, one can imagine situations in which out-degree might also play a dominant role.  Furthermore, if this model were used to model more general classes of random directed acyclic graphs, $\beta \neq 0$ might play an important role.  With $\alpha$ alone, however, we can still model a wide class of processes.  When $\alpha = 0$, we recover a model very similar to the Erdos-Renyi $G(n,m)$, as the measure becomes
\begin{align*}
	\mathbb{P}((x,y)) =& \frac{\exp(0 \delta^-(y))}{\sum_{i=1}^{|V(G_t)|} \exp(0 \delta^-_i)}\\
	=& \frac{1}{|V(G_t)|}
\end{align*}
\par
Here, each node clearly has an equal probability of receiving new arcs.  For increasing levels of $\alpha > 0$, an increasing level preferential attachment effect occurs.  When $\alpha \approx +\epsilon$, $e^{\alpha x}$ behaves practically as a linear function, and we recover linear preferential attachment.  Note, however, that unlike most linear models, even nodes with in-degree $0$ will still have some probability of citation since $e^0 = 1$.  For $\alpha >> 0$, however, we can produce a nonlinear preferential attachment mechanism.  This should be apparent from the fact that $e^{\alpha x}$ is convex increasing for $\alpha > 0, \delta^- > 0$.  Take, for example, the case in which $\delta^-(x) = 2, \delta^-(x) = 1$.  Under the B-A linear model, $x$ would be twice as likely to receive a citation.  If $\alpha = 0.5$, however, then $\frac{e^{\alpha \delta^-(x)}}{e^{\alpha \delta^-(y)}} = 1.65$, meaning $x$ is only one-and-a-half times more likely to receive citations.  If $\alpha = 1$, $\frac{e^{\alpha \delta^-(x)}}{e^{\alpha \delta^-(y)}} = e$, meaning $x$ is nearly three times as likely to receive citations.  It should thus be clear that the $\alpha$ parameter allows the same measure to produce a wide variety of specific attachment mechanisms.  It is interesting to note that one might even envision a system where high in-degree is \textit{punished}; that is, the appropriate value of $\alpha$ could be negative, and thus lower in-degree nodes would have a higher probability of receiving citations.
\par
To illustrate the effect of $\alpha$ visually, we generate realizations of the model with two values of $\alpha$. On the left, $\alpha=0$, and the model quite clearly resembles an Erdos-Renyi attachment mechanism.  On the right, $\alpha = 1$, and a strong preferential attachment mechanism is visible in the central, high in-degree node.
\begin{figure}[ht]
	\centering
	\includegraphics[width=13cm]{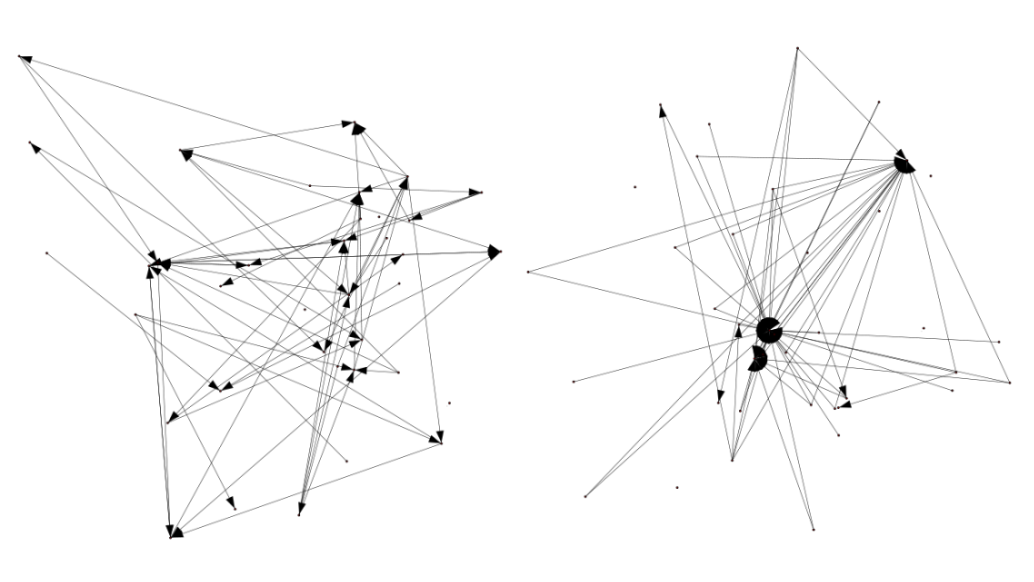}
	\caption{Example Realizations of the Model.  Left: $\alpha=0$, Right: $\alpha=1$}
\end{figure}

\par
One detail remains unspecified in this model as well as in a large amount of the literature involving the simulation of these graphs.  This detail, however, has a profound influence on the resulting graph structure, and hinges on the following question:  are arcs for a new node sampled from $\Omega_t^A$ with or without replacement?  If the arcs are sampled with replacement, the resulting graph is either an arc-weighted digraph or a directed multigraph.  If the arcs are sampled without replacement, then the graph is guaranteed to have no multiple arcs, and thus remains a simple digraph.  These distinctions obviously result in very different models and lead to different appropriate algorithms.  Most prior literature silently chooses sampling without replacement, and we will follow this choice.  

With the model fully specified, we now proceed to derive a number of properties.  We want to investigate the generation of sinks, whose importance is discussed above.  If $x$ is a node, then the probability that $x$ is a sink is given by
\begin{align*}
	\mathbb{P}(\delta^+(x) = 0) =& \frac{e^{-\mu} \mu^0}{0!}\\
	=& e^{-\mu}
\end{align*}
where $\delta^+(x)$ is the out-degree of $x$ and $\frac{e^{-\lambda} \lambda^k}{k!}$ is the probability density function for a Poisson distribution.
\par
Let $S_i$ be the Bernoulli trial representing whether or not the $i^{th}$ node is a sink.  Then we can write the number of sinks at time $t$ as
\begin{align*}
	S(t) =& \sum_{i=1}^{|V(G_t)|}  S_i
\end{align*}
\par
Since $|V(G_t)|$ is a Poisson process and the $S_i$ are IID random variables, $S(t)$ is a compound Poisson process (Grimmet and Stirzaker 2001 \cite{Grimmet2001}).  By the properties of compound Poisson processes, we know that
\begin{align*}
	\mathbb{E}[S(t)] =& \mathbb{E}[|V(G_t)| S_i]\\
	=& \mathbb{E}[|V(G_t)|] \mathbb{E}[S_i]\\
	=& \lambda t e^{-\mu}
\end{align*}
\par

The next property we would like to investigate is density.  Density represents how many arcs exist in a graph relative to how many arcs \textit{could have} existed.  In general, the density of a simple digraph is given by
\begin{align*}
	D =& \frac{|A|}{2 {|V| \choose 2}}\\
	 =& \frac{|A|}{|V| (|V| - 1)}
\end{align*}
where $|A|$ is the number of edges and $|V|$ is the number of vertices of an undirected graph.
\par
However, since citation networks are directed acyclic graphs, there exists a topological ordering of the vertices such that the corresponding adjacency matrix is triangular.  This is equivalent to the condition above that $T(x) \leq  T(y) \Leftrightarrow \mathbb{P}((x,y) \in A(G)) = 0$.  Therefore, the equation for the density of a directed acyclic graph is the same as that of an undirected graph: 
\begin{align*}
	D =& \frac{|A|}{{|V| \choose 2}}\\
	  =& \frac{2 |A|}{|V| (|V| - 1)}
\end{align*}
where $|A|$ is the number of arcs and $|V|$ is the number of vertices of a directed graph.  Since $|V|$ increases with $\lambda$, and $|A|$ increases with $\lambda$ and $\mu$, we see that the density of the graph can be controlled with these parameters.  

\section{Results}
In order to determine the suitability of the standard community
detection algorithms, we generate random graphs for various values of $\lambda,
\mu, $ and $\alpha$.  Varying these parameters allows us to produce random
graphs with specific characteristics.  Although this model is somewhat stylized,
these controlled characterizations provided in the following simulations allow
us to ultimately provide prescriptions to researchers who would like to choose
between these algorithms based on quantitative or qualitative understandings of
their own data.
\par
For each of the following model configurations, we produce 50 random
realizations of the graph based on the specified parameters, where each
realization is comprised of 200 time steps of the discrete dynamics
\footnote{Varying the number of steps taken by the model will almost certainly
produce different results.  However, as the point of this paper is investigate
differences between algorithms for different model parameters and not on
different time scales, we leave this investigation for future work.}.  To
reiterate the details of section 3, $\lambda$ controls the Poisson rate
of nodes per time step, $\mu$ controls the Poisson rate of arcs per new node,
and $\alpha$ controls the equation that determines the probability of attachment
to existing nodes.  The initial graph is comprised of a single node and arcs are
sampled without replacement.
\par
For each of these realizations, the community structure of the graph
is calculated at each step of the model and stored.  To calculate this
community structure, we choose to compare the three most commonly applied
community detection algorithms: Girvan-Newman edge-betweenness (Girvan and
Newman 2002\cite{Girvan2002}), fast-greedy (Newman 2004 \cite{Newman2004}), and
leading eigenvector (Newman 2006\cite{Newman2006}).  For the edge-betweenness
algorithm, a directed implementation is used.  However, for both the fast-greedy
and leading eigenvector algorithms, undirected implementations are used.  As
these algorithms can be sensitive to implementation details, it is important to
note that the software package chosen for this comparison is igraph version 0.6
(Csardi and Nepusz 2006 \cite{Csardi2006}). igraph is one of the most widely
used packages for community detection, and thus running our calculations with
the algorithms provided by this package produces results relevant to most
researchers.  Furthermore, though directed versions of both fast-greedy and
leading eigenvector are possible, their use is impractical for the majority of
researchers since their implementations are not publicly available (Leicht and
Newman 2008 \cite{Leicht2008}).
\par
Thus, each realization produces a time-indexed set of community memberships for
each algorithm.  From this data, we calculate two primary statistics: a
pairwise community stability measure and the average number of communities
detected across time steps per realization.
\par Measuring community stability is an idea introduced by Palla, et al. 2007
\cite{Palla2007}.  Their conception of stability, however, has a number of
subtleties that may complicate interpretation of the result.  We
introduce a pairwise community stability measure that captures the likelihood
that pairs of nodes in the same community will remain so in future time
steps . More explicitly, pairwise community stability can be expressed
probabilistically as
\begin{align}
	\mathbb{E}[\mathbb{P}(C_i(t) = C_j(t) | C_i(t-1) = C_j(t-1))]
\end{align}
where $C_i(t)$ is the community ID of the $i^{th}$ node at time $t$.  The
``appropriate'' range of stability depends on the amount of expected microscopic
fluctuation in the network.  If a system is primarily comprised of dyadic
relationships that are fairly stable, then pairwise stability should be fairly
high.  On the other hand, if only macroscopic structures are stable or no
structure exists, then pairwise stability should be low.  In the context of
citation networks, however, we expect that two documents should typically remain
in the same community after being placed together.  Thus, though other
modifications to the Palla stability conception are possible, pairwise stability
provides a more intuitive interpretation for some contexts (Fenn, et
al. 2009 \cite{Fenn2009}).
\par
The second statistic we calculate is the average number of communities across
time steps per realization.  This is a measure of the granularity or level
of detail represented by an algorithm's clustering. An algorithm that
produces many more communities than another algorithm for the same graph must
necessarily produce communities that are, on average, smaller in size.  This can
be interpreted as focusing on a different scope of structure, but stability
might vary across different scopes of the graph.  This quantity can be written
as
\begin{align}
	\frac{1}{T	} \sum_{t=1}^T \max_i C_i(t)
\end{align}
\par
The first model configuration chosen is the simplest possible: a $\lambda$ rate of one node per step, a $\mu$ rate of one arc per node, and an Erdos-Renyi $\alpha$ parameter of $0$.  As applied to a citation network, this is equivalent to tracking a field in which there is on average one paper per step with one citation per paper.  While this is a highly stylized baseline case, we offer it as the simplest citation process possible.  Figure 2 belows shows that even in this simplest case, there is a stark difference between these algorithms.  The upper left pane shows the distribution of pairwise community stability across realizations of the model for each algorithm.  The upper right pane shows the distribution of the average number of communities detected across realizations of the model for each algorithm.    
\begin{figure}[h!]
	\includegraphics[width=6.5cm]{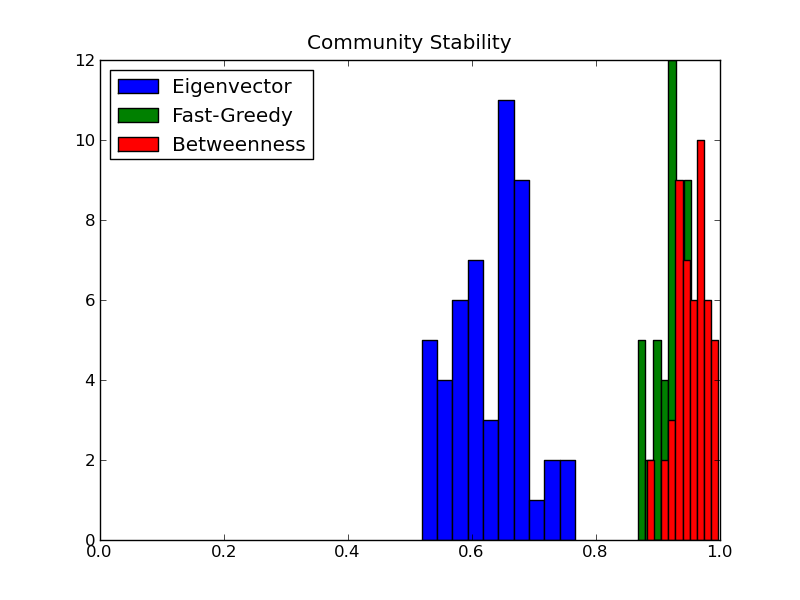}
	\includegraphics[width=6.5cm]{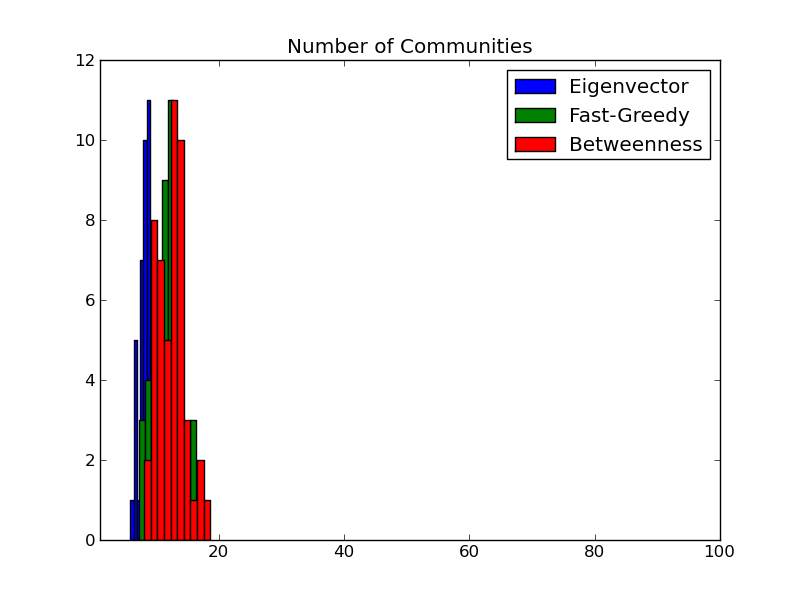}
	\caption{Left: Community Stability.  Right: \# of Communities.}
\end{figure}

Below is a table of the means and standard deviations of each algorithm for both panels of Figure 2.
\begin{center}
\begin{tabular}{c|c|c|c}
	\tiny
	& \textrm{Eigenvector} & \textrm{Fast-Greedy} & \textrm{Betweenness}\\
	\hline
	\textrm{Stability: Mean(Std)} & 0.63 (0.061) & 0.93 (0.029) & 0.95 (0.026)\\
	\hline
	\textrm{\# Communities: Mean(Std)} & 8.5 (1.09) & 11.4 (2.06) & 12.3 (2.18)
	\normalsize
\end{tabular}
\end{center}
\vspace{1cm}
\par
An even more informative representation of these data is a scatter plot representing the stability value and the number of communities detected for each realization.  The slope of each algorithm's data points and their positions relative to each other in the scatter plot communicate a great deal about the tradeoffs between these algorithms.  Figure 3 below shows that for this model configuration, the eigenvector method is producing a lower variation in number of communities at the expense of a lower stability.  Both fast-greedy and betweenness produce much higher stability values by dividing the network into smaller communities.
\par
\begin{figure}[h!]
	\includegraphics[width=13cm]{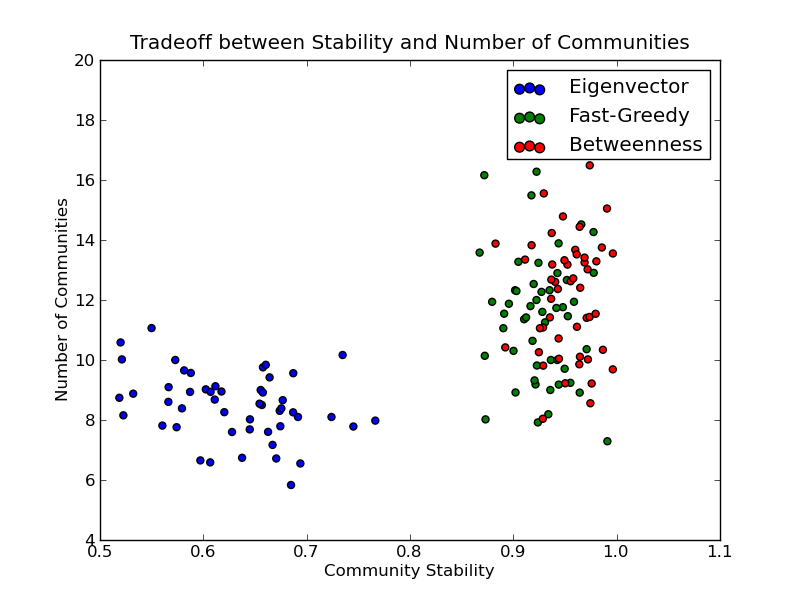}
	\caption{Scatter of stability and \# of communities.}
\end{figure}
For this configuration, varying $\alpha$ had little to no effect on these algorithm measures.  
\par
For our next experiment, we increase the average number of arcs per node $\mu$ to 4 while holding $\lambda=1$ and $\alpha=0$.  Increasing this ratio of $\frac{\mu}{\lambda}$ results in a higher graph density, and thus it is more likely that the graph is more strongly connected.  Figure 4 below again shows the scatter plot of stability and number of communities.  This model configuration produces significantly different results relative to the previous experiment.  Now, both eigenvector and fast-greedy settle on many fewer communities than betweenness.  Despite this, fast-greedy produces a higher stability than eigenvector.  Betweenness and fast-greedy have equivalent stabilities, despite nearly an order-of-magnitude difference in the number of communities detected.  It is interesting to note that eigenvector and fast-greedy, both modularity-based approaches, appear to fall along a single curve in this space.
\par
\begin{center}
\begin{tabular}{c|c|c|c}
	\tiny
	& \textrm{Eigenvector} & \textrm{Fast-Greedy} & \textrm{Betweenness}\\
	\hline
	\textrm{Stability: Mean(Std)} & 0.70 (0.044) & 0.80 (0.044) & 0.80 (0.43)\\
	\hline
	\textrm{\# Communities: Mean(Std)} & 6.90 (0.85) & 4.76 (0.39) & 13.70 (2.73)
	\normalsize
\end{tabular}
\end{center}
\vspace{1cm}

\begin{figure}[h!]
	\includegraphics[width=13cm]{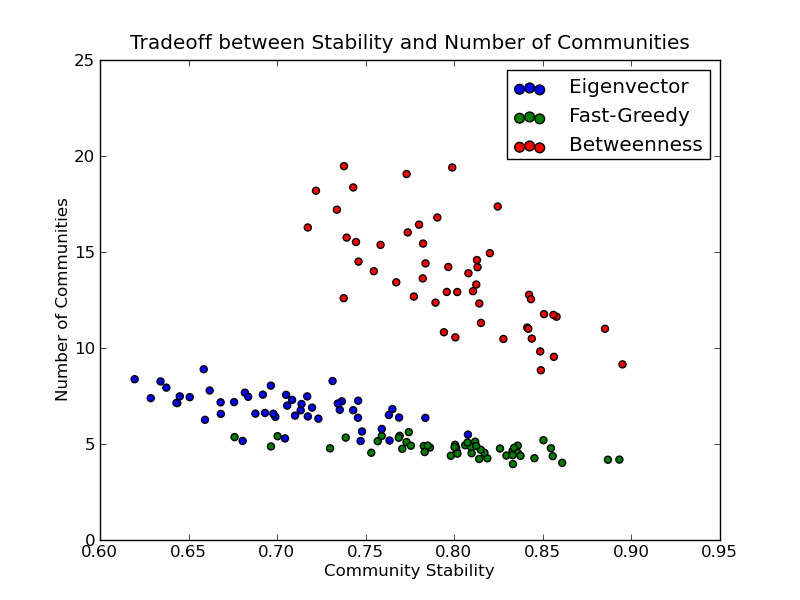}
	\caption{Scatter of stability and \# of communities.}
\end{figure}

The above results are calculated for the experiment with $\alpha = 0$, however.  We are also interested in whether or not a preferential attachment mechanism has a significant effect on the suitability of these algorithms.  To do so, we left the model parameters $\lambda = 1$ and $\mu = 4$ but increased $\alpha$ to 1.  As discussed in Section 3, an $\alpha$ value of 1 produces a very strong preferential attachment mechanism in which the highest in-degree nodes are most likely to receive new arcs.  Figure 5 below shows the scatter plot for this configuration.  As both the table and the figure indicate, the performance of betweenness has been altered dramatically.  Not only has its stability increased significantly, but the variance of its stability has dropped as well.  Eigenvector and fast-greedy remain relatively unchanged and remain along the same curve; interestingly, betweenness appears to have collapsed onto this curve as well.  These results indicate that an analytic functional form may exist to represent the tradeoff between stability and number of communities for certain classes of graphs.
\par
\begin{center}
\begin{tabular}{c|c|c|c}
	\tiny
	& \textrm{Eigenvector} & \textrm{Fast-Greedy} & \textrm{Betweenness}\\
	\hline
	\textrm{Stability: Mean(Std)} & 0.74 (0.062) & 0.87 (0.32) & 0.95 (0.020)\\
	\hline
	\textrm{\# Communities: Mean(Std)} & 7.62 (1.78) & 4.31 (0.60) & 4.98 (1.33)
	\normalsize
\end{tabular}
\end{center}
\vspace{1cm}

\begin{figure}[h!]
	\includegraphics[width=13cm]{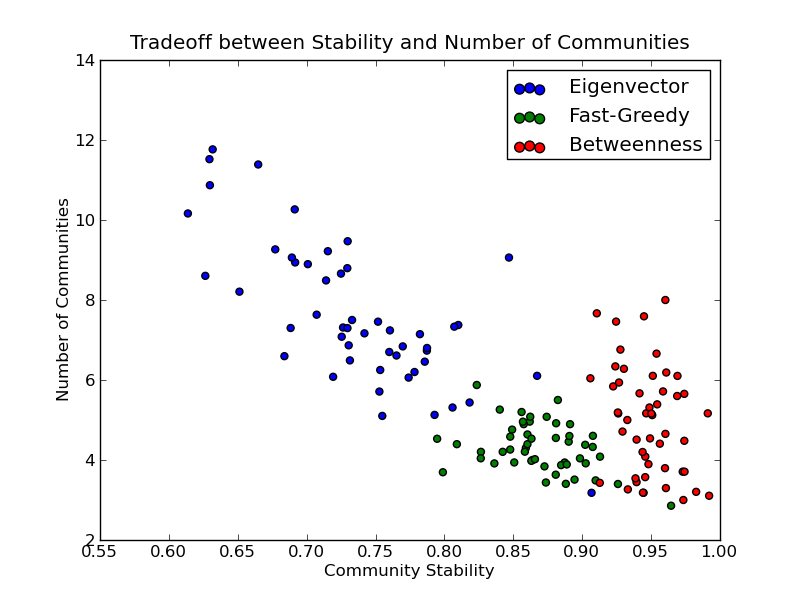}
	\caption{Scatter of stability and \# of communities.}
\end{figure}
\pagebreak

The previous experiment models a more dense, slowly growing citation process.  In the following experiment, we switch the values of $\lambda$ and $\mu$ to investigate these algorithms on less dense, quickly growing citation network.  Now, $\lambda = 4$, $\mu = 1$, and we initially set $\alpha = 0$.  The results in the following table and Figure 6 show a relationship very similar to the previous experiment with $\alpha=1$ in Figure 5.  However, though the stability ordering is preserved, betweenness now yields a higher stability by producing more communities, not fewer.  This result indicates that the slope of the above-mentioned functional form likely depends on the ratio $\frac{\mu}{\lambda}$ and the resulting density of the graph.
\begin{center}
\begin{tabular}{c|c|c|c}
	\tiny
	& \textrm{Eigenvector} & \textrm{Fast-Greedy} & \textrm{Betweenness}\\
	\hline
	\textrm{Stability: Mean(Std)} & 0.61 (0.038) & 0.75 (0.057) & 0.91 (0.020)\\
	\hline
	\textrm{\# Communities: Mean(Std)} & 24.27 (2.00) & 31.13 (3.55) & 39.85 (4.74)
	\normalsize
\end{tabular}
\end{center}
\vspace{1cm}

\begin{figure}[h!]
	\includegraphics[width=13cm]{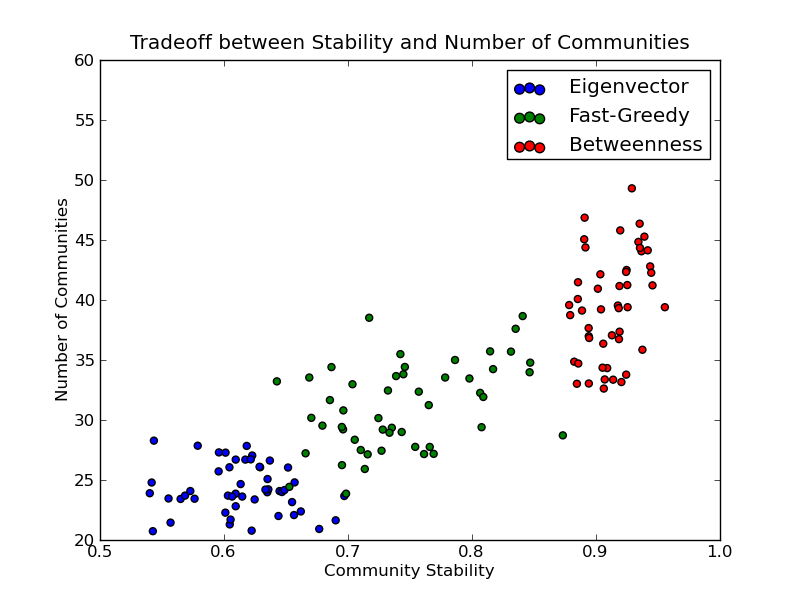}
	\caption{Scatter of stability and \# of communities.}
\end{figure}

\pagebreak
Again, these results incorporate no preferential attachment.  To observe the effect of preferential attachment, we increase $\alpha$ to 1 and observe the outcomes.  Figure 7 and the below table show the results.  With $\alpha=1$ for this quickly growing, less dense network, fast-greedy and betweenness are statistically equivalent.  However, the continuous gradient from eigenvector to fast-greedy has been broken, with a large difference between the two algorithms in terms of both stability and number of communities.  
\begin{center}
\begin{tabular}{c|c|c|c}
	\tiny
	& \textrm{Eigenvector} & \textrm{Fast-Greedy} & \textrm{Betweenness}\\
	\hline
	\textrm{Stability: Mean(Std)} & 0.72 (0.049) & 0.96 (0.034) & 0.96 (0.020)\\
	\hline
	\textrm{\# Communities: Mean(Std)} & 18.76 (4.35) & 41.74 (5.35) & 47.99 (7.41)
	\normalsize
\end{tabular}
\end{center}
\vspace{1cm}

\begin{figure}[h!]
	\includegraphics[width=13cm]{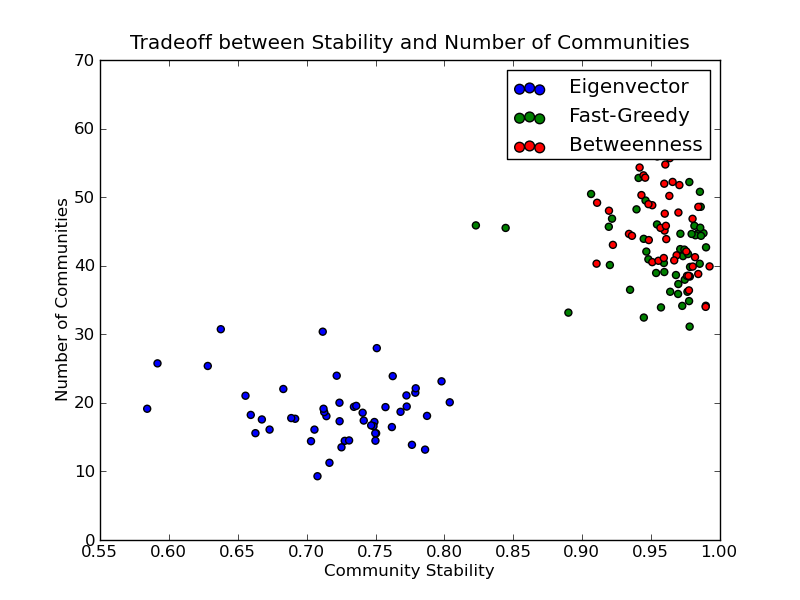}
	\caption{Scatter of stability and \# of communities.}
\end{figure}

\pagebreak
\section{Future Work \& Conclusion}
In this paper, we develop a novel analytic and computational model for random
directed acyclic digraphs.  Our model is capable of representing a wide
class of systems and produces results demonstrating significant differences in
the output of three canonical community detection algorithms.  While these
results clearly counsel against a one-size-fits-all approach to community
detection, they also strongly suggest broad classes of network problems
exist for which existing community detection algorithms are applicable. We
believe that this is the first step towards the creation a toolbox of diagnostic
heuristics and techniques that will be useful for working social scientists who
want to perform community detection with confidence in the stability and
accuracy of the algorithms available to them.

\pagebreak
\section{Acknowledgments}
We would like to thank the University of Michigan Center for the Study of
Complex Systems (CSCS) for providing both computational resources and a
fruitful research environment.  We would also like to thank Peter J. Mucha,
Mason A. Porter, and James H. Fowler for helpful comments, contributions, and/or
assistance.

\end{document}